\documentclass[onecolumn, preprintnumbers, showpacs, nofootinbib, aps]{revtex4-1}
\usepackage{graphicx}

\newcommand{\bi}{\bibitem}
\newcommand{\be}{\begin{eqnarray}}
\newcommand{\ee}{\end{eqnarray}}

\begin{document}

\title{Thick disk accretion in Kerr space-time with arbitrary spin parameters}

\author{Cosimo Bambi}
\email{cosimo.bambi@ipmu.jp}

\author{Naoki Yoshida}
\email{naoki.yoshida@ipmu.jp}

\affiliation{
Institute for the Physics and Mathematics of the Universe, 
The University of Tokyo, Kashiwa, Chiba 277-8583, Japan}

\date{\today}

\preprint{IPMU10-0147}

\begin{abstract}
In this paper we extend our previous works on spherically symmetric
accretion onto black holes and super-spinars to the case in which
the fluid has a finite angular momentum initially. We run 2.5D and 3D
general relativistic hydrodynamic simulations of
the accretion of a fat disk. We study how the accretion process changes
by changing the values of the parameters of our model. We show that
the value of the fluid angular momentum critically
determines turn-on and off the production of powerful equatorial 
outflows around super-spinars. For corotating disks, equatorial outflows 
are efficiently generated, 
even for relatively low spin parameters or relatively large 
super-spinar radii. For counterrotating disks, equatorial outflows 
are instead significantly suppressed, and they are possible only 
in limited cases. We also study accretion around a tilted disk.
\end{abstract}

\pacs{04.20.Dw, 97.60.-s, 95.30.Lz, 97.10.Gz}

\maketitle


\section{Introduction}

Today we believe that the final product of the gravitational collapse 
is a Kerr black hole. Nevertheless, observational evidences supporting
this conjecture are still weak~\cite{narayan}. Astronomical
observations have led to the discovery of at least two classes of
astrophysical black hole candidates\footnote{The existence of a third 
class of astrophysical black holes, intermediate objects with mass 
$M \sim 10^2 - 10^4$~$M_\odot$, is still controversial, because 
dynamical measurements of their masses are uncertain.}: 
stellar-mass objects in X-ray binary systems ($M \sim 5 - 20$~$M_\odot$)
and super-massive objects at the center of most galaxies 
($M \sim 10^5 - 10^{10}$~$M_\odot$). All these objects are supposed to 
be Kerr black holes because they cannot be explained by something 
else without introducing new physics. For example, stellar-mass black 
hole candidates in X-ray binary systems are too heavy to be neutron/quark
stars for any reasonable matter equation of state~\cite{ruf,kal}. 
The super-massive black hole candidate at the center of the Galaxy 
is too massive, compact, and old to be a cluster of non-luminous 
bodies~\cite{maoz}.

Gravity is relatively well tested only in the weak field limit~\cite{will}; 
we do not know if our theory is still reliable in the case of strong
gravitational fields. Indeed, the idea that the 
final product of the gravitational collapse is a Kerr black hole 
itself is based on a set of assumptions, which may be wrong, see e.g. 
the discussion in~\cite{talk} and references therein. 
New physics may not be so unlikely. There are 
a couple of fundamental problems associated with the existence of black 
holes when quantum effects are taken into account; several authors
have thus suggested other possibilities as the final product of the
collapsing matter~\cite{mottola}. Lastly, quantum gravity effects 
may become important already at the gravitational radius of a system, 
$R_g = M$, rather than at the Planck scale 
$L_{Pl} = 10^{-33}$~cm~\cite{samir}; if this is true, astrophysical 
black holes might be quite different objects from the ones predicted 
in the classical theory.

The Kerr metric has only two free parameters, associated respectively 
with the mass, $M$, and the spin, $J$, of the massive object. The
condition for the existence of the event horizon is $|a_*| \le 1$,
where $a_* = J/M^2$ is the dimensionless spin parameter. It is also
remarkable that, at least in some circumstances, the accretion process
should spin the black hole up to $a_* \approx 0.998$~\cite{thorne}.
For $|a_*| > 1$, there is no black hole solution, but the Kerr 
metric may still be used to describe the vacuum gravitational 
field around some very fast rotating body.
If current black hole candidates are not Kerr black holes but, say, 
compact bodies made of some kind of exotic matter, the maximum value 
of the spin parameter may be either larger or smaller than 1. The
determination of the spin parameter can thus be used to test the 
nature of these objects. If we find that one of them violates the 
bound $|a_*| \le 1$, then the final product of the gravitational 
collapse is not (or not necessarily) a Kerr black hole. The possible
discovery that all these objects have spin parameter significantly 
smaller than 1 may suggest the same conclusion. The possibility of 
the existence of compact objects with $|a_*| > 1$ (super-spinars) 
was first discussed in~\cite{gim-hor}. Some basic features of the
electromagnetic radiation emitted around a super-spinar were studied 
in~\cite{pap1,pap2,rohta}, and compared with the one emitted from 
a Kerr black hole.

Most of the observable astrophysical phenomena occurring around a
compact object are significantly determined by the exact accretion
process~\cite{review}. General relativistic effects are important
only very close to the massive object, where gravity is stronger.
The special case of spherically symmetric and
adiabatic accretion onto a Schwarzschild black hole can be studied
analytically~\cite{michel}. In general, instead, a numerical
approach is necessary. The first numerical simulations in Kerr
background were presented in~\cite{wilson} and then in~\cite{hsw1,hsw2}.
After these works, the research on the accretion process in
Kerr space-time was devoted largely to the study of black hole tori and of
the associated instabilities~\cite{hawley,i-b,devilliers}. More
recent works have investigated the role of magnetic fields and
the mechanisms to generate jets; for a review, see e.g. Ref.~\cite{font}.
In~\cite{sim1,sim2,sim3}, we extended the study of
spherically symmetric and adiabatic accretion onto Kerr black holes 
to the case of spin parameter $|a_*| > 1$. The most important result
was the discovery of strong equatorial
outflows by the sole gravitational force from fast-rotating 
super-spinars.

In this paper, we study the accretion process onto black holes and
super-spinars when the angular momentum of the gas is
initially non-negligible. We study the case of a geometrically thick disk.
In our previous works, the accretion process depended on the value
of the spin parameter and on the radius of the super-spinars,
while other conditions, such as the gas equation of state, had a minor
role. Here we show that even the gas angular momentum is
very important in the accretion process. 
Spin-orbit interactions can indeed strongly foster the production
of equatorial outflows in the case of corotating disks, while suppress
it for counterrotating disks. This means that
equatorial outflows can be produced even for relatively low spin
parameters and/or relatively large super-spinar radii.

The paper is organized as follows. In Sec.~\ref{s-bondi}, we briefly
review the basic features of spherically symmetric and adiabatic
accretion onto black holes and super-spinars. In Sec.~\ref{s-disk},
we present our new hydrodynamics simulations of thick
disk accretion and we discuss our findings. In Sec.~\ref{s-conc},
we report our conclusions. Throughout the paper we use 
Boyer-Lindquist coordinates to describe the Kerr background and 
natural units $G_N = c = k_B = 1$.

\section{Spherically symmetric accretion \label{s-bondi}}

Spherically symmetric and adiabatic accretion in Kerr space-time
with arbitrary value of the spin parameter $a_*$ was studied in
Refs.~\cite{sim1,sim2,sim3}, in 2.5 and 3 dimensions. A summary
of the numerical approach and of the main results can be found
in~\cite{talk}. The behavior of the accretion process is essentially
determined by $a_*$ and by the physical radius of the compact object, 
$R$. Because of the ergoregion instability, it is likely that 
super-spinars cannot be extremely compact with $R \ll M$~\cite{enrico}. 
However, they can be stable if their physical radius is of the 
same order of their gravitational radius and, in~\cite{sim3}, we 
assumed $R = 2.5$~$M$. The accretion process onto a super-spinar with
smaller/larger physical radius can be easily recovered from the case
with $R = 2.5$~$M$ by considering a super-spinar with larger/smaller
spin parameter.

There are three qualitatively different kinds of accretion:
\begin{enumerate}
\item {\it Black hole accretion}. For black holes and super-spinars 
with $|a_*|$ moderately larger than 1, one finds the usual accretion 
process onto a compact object. The increase in $|a_*|$ makes the 
accretion process more difficult: in the quasi-steady-state 
configuration, the velocity of the gas around the compact object is 
lower, while the density and the temperature are higher. The 
gravitational field indeed becomes weaker for higher spin parameters. 
One can easily understand this by noticing that the radius of the event
horizon of a black hole monotonically decreases with $a_*$. The 
difference, however, is very small and the exact value of the spin
parameter does not affect significantly the process. 
\item {\it Intermediate accretion}. As the spin parameter increases, 
the gravitational force around the super-spinar becomes weaker and,
at some point, can become repulsive. 
Now the accretion process is significantly suppressed: 
the flow around the super-spinar becomes subsonic and the density and 
the temperature of the gas increase further. 
\item {\it Super-spinar accretion}. For high values of the spin parameter,
the process of accretion is quite different: matter is accreted mostly from the 
poles, while the repulsive gravitational field produces outflows around 
the equatorial plane. For $R = 2.5$~$M$, this occurs for $|a_*| \ge 2.9$,
while for smaller/larger values of $R$, powerful equatorial outflows
appear for lower/higher values of $|a_*|$.
\end{enumerate}

The outflows generated around super-spinars have no counterpart
in the case of black holes. When $|a_*| \le 1$, jets and outflows
cannot be produced in spherically symmetric and adiabatic accretion. 
The most popular scenarios to generate jets around astrophysical
black holes require the presence of magnetic fields 
({\it magnetically-driven outflows}).
For fast-rotating black holes, the Blandford-Znajek mechanism~\cite{b-z}
seems to be the most promising and efficient way to extract energy 
from the black hole and produce relativistic jets.
For slow-rotating or Schwarzschild black holes, jets may instead 
be powered by the Blandford-Payne mechanism~\cite{b-p}.
Current studies, however, are controversial, and there is not
a common agreement on the efficiency of these processes, see
e.g. Refs.~\cite{mckinney} and \cite{livio}. If the mass accretion 
rate is close or exceeds the Eddington limit, quasi-steady outflows 
with high velocity ($v \sim 0.25$) can also be driven by 
the force of the radiation pressure
({\it radiation driven outflows})~\cite{blinnikov,ohsuga}. 
A model in which jets are driven by the force of the radiation
pressure and collimated by the magnetic field has been recently
proposed in~\cite{ohsuga2}.
In all these models, matter should 
be ejected parallel to the spin of the massive object and/or 
perpendicular to the accretion disk\footnote{In the case of stellar-mass
black holes in X-ray binary systems, one can expect that the spin 
of the black hole is usually perpendicular to the accretion disk,
on the basis of binary population synthesis~\cite{fragos}. For
super-massive black holes, at least the central part of the 
accretion disk may lie on the equatorial plane of the black hole
because of the Bardeen-Petterson effect~\cite{b-pet}. However,
there are observational data~\cite{maccarone} and theoretical 
arguments~\cite{fragile} suggesting that tilted disks may instead 
be common.}.

The outflows found in our hydrodynamics simulations in the case of
super-spinars are instead {\it gravitationally-driven outflows}:
they are produced by the peculiar features of the
gravitational field in the Kerr metric when $|a_*| > 1$. These
outflows can thus be generated even when the accretion process is
adiabatic and (at large radii) spherically symmetric. Here
matter is ejected on and around the equatorial plane; that is,
perpendicular to the spin of the massive object. In some circumstances, 
the amount of matter in the outflow is considerable, which can
significantly reduce the mass accretion rate onto the central
object.

\section{Thick disk accretion \label{s-disk}}

\begin{figure}
\par
\begin{center}
\includegraphics[height=6cm,angle=0]{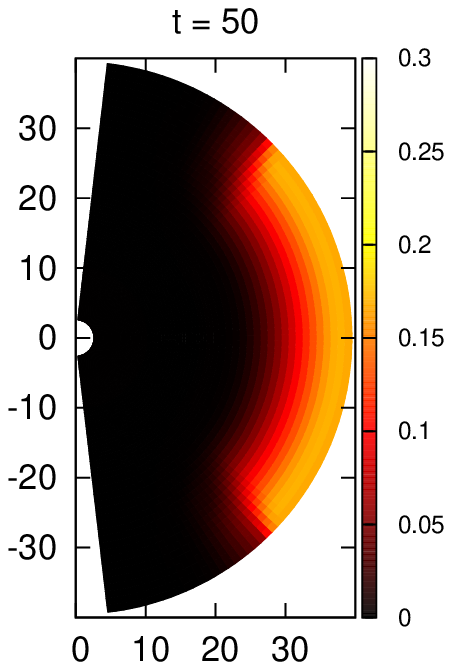}
\includegraphics[height=6cm,angle=0]{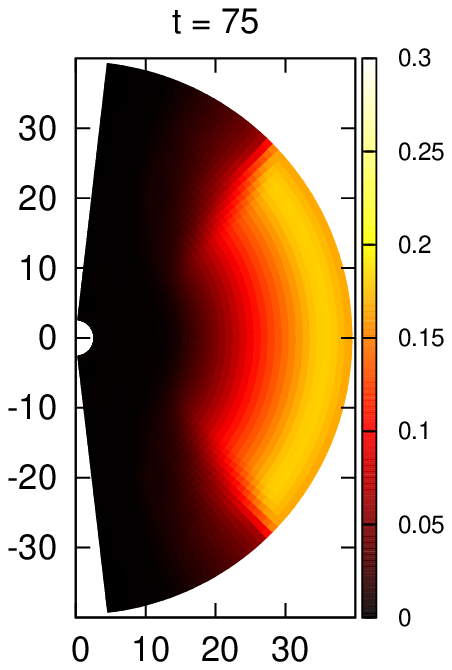}
\includegraphics[height=6cm,angle=0]{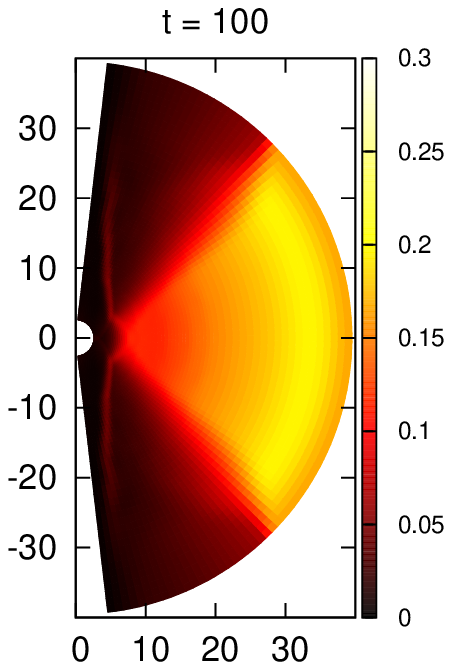}
\end{center}
\par
\vspace{-5mm} 
\caption{Snapshots of the gas density (in arbitrary 
units) around a Schwarzschild black hole
at $t = 50$~$M$ (left panel), $t = 75$~$M$ (central panel), 
and $t=100$~$M$ (right panel), respectively.
The unit of length along the axes is $M$.}
\label{f-snapshots}
\end{figure}

To study the thick disk accretion, we use the same set-up adopted in
Ref.~\cite{sim3}. The accreting matter is modeled as an ideal
fluid with polytropic index $\Gamma = 5/3$. Its gravitational
field is neglected. Mass increase and the variation
in spin of the massive object, as a consequence of the accretion
process are also neglected. 
Our study is based on simulations in 2.5 dimensions,
and the computational domain is $2.5 \, M < r < 40 \, M$ and 
$0 < \theta < \pi$. So, we are interested only in the innermost 
part of the accretion disk, where the properties of the gravitational
field are more important. We run a few simulations in full three dimensions 
in order to check that our main conclusions based 2.5 dimensions 
are robust. 
The initial configuration and the boundary conditions are
different from Ref.~\cite{sim3}. Here, the computational domain 
is initially empty. We inject gas from the outer boundary
for $\pi/4 < \theta < 3\pi/4$. For $\theta < \pi/4$ and 
$\theta > 3\pi/4$, we impose free-outflow boundary conditions, 
i.e. we set zero gradient across the boundary: this condition
allows for both outflow and inflow and it is a quite common
choice for this kind of simulations~\cite{i-b}. 
The gas is injected with
non-zero radial and azimuthal velocity. Our default choice at the
beginning of the simulations is $v^{\rm r} = -0.15$ and 
$v^{\phi} = 0.0010/\sin\theta$,
corresponding to a gas angular momentum per unit mass $l_0 = 1.6$~$M$.
The polar velocity, $v^\theta$, is set to be zero. Snapshots of the
gas density at $t = 50$~$M$, 75~$M$ and 100~$M$,
are shown in Fig.~\ref{f-snapshots}, for the case of $a_* = 0$.

\subsection{Black holes}

We first discuss the case $a_* = 0$. The accretion process 
is determined by the gas angular momentum, which is controlled by 
the injection azimuthal velocity at the outer boundary. For low 
specific angular momenta, the gas reaches the inner boundary and 
is swallowed by the black hole. As the angular momentum increases, 
the height of the effective potential barrier increases as well.
At some point, the gas cannot reach the central object and there is 
not accretion any more. The critical value of the specific angular 
momentum depends on the accretion rate and on the mechanism 
responsible for transporting angular momentum to larger radii. 
If the infall gas velocity is too low, the gas injected from the outer
boundary clashes with the one already present in the computational
domain, generating turbulences and random motions, a part of which 
leaves the computational domain. In this case, the gas has not the 
time to lose angular momentum to fall to the central object. 
On the other hand, if angular momentum 
transport by turbulent motions is efficient, 
the gas can reach quickly the black hole and new gas can enter easily
into the computational domain. 
For higher specific angular momenta, the 
gas is confined at larger distance from the black hole. In all
these simulations, the exact value of the injection radial velocity 
is not very important. For higher $v^r$, the gas reaches 
the central object somewhat quicker, 
but for reasonable choices of its value we 
do not see a significant difference.

\begin{figure}
\par
\begin{center}
\includegraphics[height=6cm,angle=0]{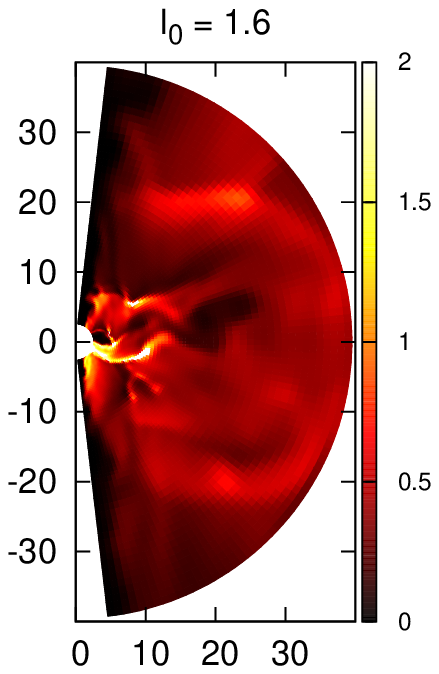}
\includegraphics[height=6cm,angle=0]{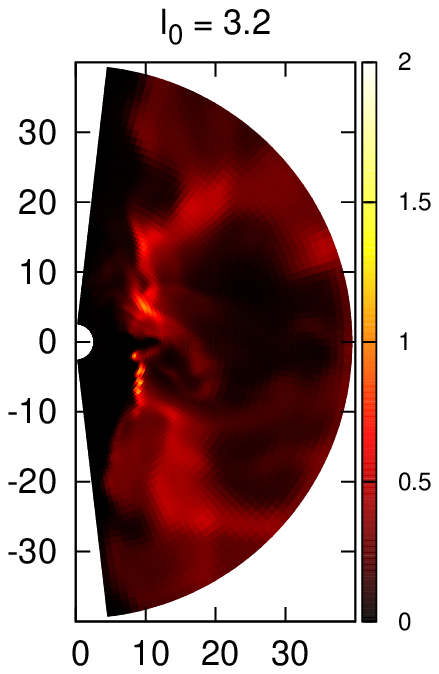}
\includegraphics[height=6cm,angle=0]{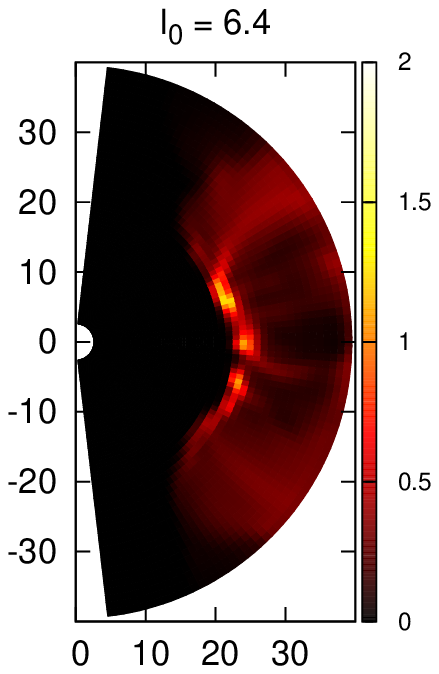}
\end{center}
\par
\vspace{-5mm} 
\caption{Snapshot at $t = 1000$~$M$ of the gas density (in arbitrary 
units) around a Schwarzschild black hole for three different values
of the initial specific angular momentum $l_0$. For $l_0 = 1.6$ 
there is accretion; in the other cases there is no accretion. In the
case of higher specific angular momentum, $l_0 = 6.4$, the gas is
confined far from the black hole. The unit of length along the axes 
is $M$; $l_0$ in unit $M = 1$.}
\label{f-den}
\end{figure}

\begin{figure}
\par
\begin{center}
\includegraphics[height=9cm,angle=0]{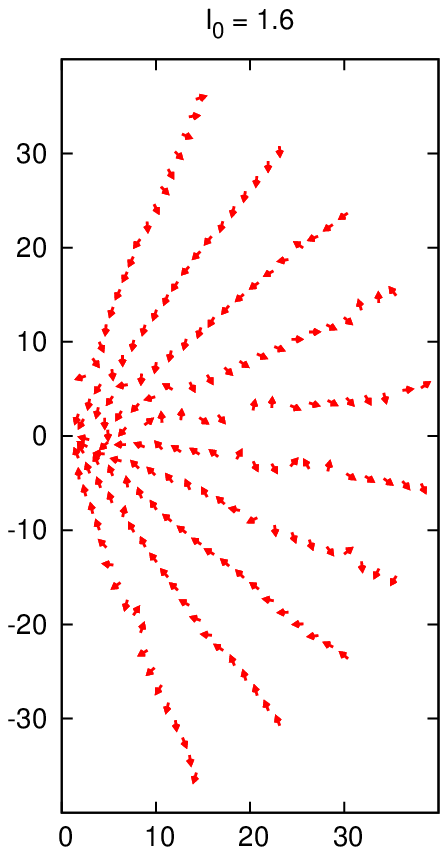} \hspace{.3cm}
\includegraphics[height=9cm,angle=0]{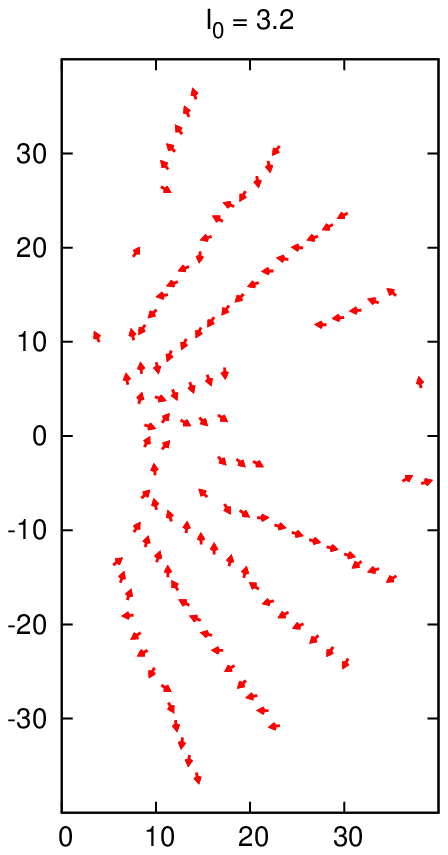} \hspace{.3cm}
\includegraphics[height=9cm,angle=0]{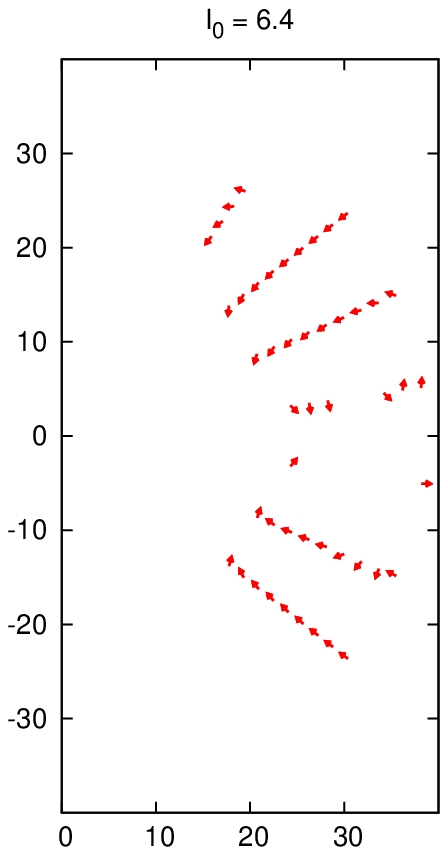}
\end{center}
\par
\vspace{-5mm} 
\caption{Directional velocity of the gas in the three cases 
presented in Fig.~\ref{f-den}. Only the region of the computational
domain with density $\rho > 0.1$ is considered.
The unit of length along the axes is $M$.}
\label{f-vec}
\end{figure}

The density of the gas around a Schwarzschild black hole is shown
in Fig.~\ref{f-den} for three different values of the initial
specific angular momentum, $l_0$. The exact value of $l_0$, however, 
should be taken with caution. It is model-dependent, and also 
it specifies only the specific angular momentum of the gas injected 
from the outer boundary at the beginning of the simulations. 
As the system
evolves, such a quantity changes its value according to the density
and the pressure of the gas which is already inside the computational 
domain.
If it were not so, we would obtain an unphysical gas configuration.
In practice, the value of $l_0$ indicates the initial degree of
gas rotation. 
The corresponding direction of the velocity of 
the gas is shown in Fig.~\ref{f-vec}. In these picture, we clearly 
see that accretion from a thick disk is characterized by chaotic
outflows, a feature absent in the case of negligible angular 
momentum of the gas. The existence of outflows resulting from fluid
rebounding at the centrifugal barrier is a well known phenomenon
in the case of accretion from thick disk and was already found 
in~\cite{wilson, hsw2}.

A reasonable estimate of the critical value of the specific angular
momentum separating the case of accretion from the one of non-accretion
onto a Schwarzschild black hole can be obtained by considering 
the motion of a test-particle in the Schwarzschild space-time. The
effective potential for a test-particle with specific angular momentum 
$l$ is
\be
V_{\rm eff} = - \frac{M}{r} + \frac{l^2}{2r^2} - \frac{M l^2}{r^3} \, .
\ee
The shape of the potential depends on the value of $l$. For
$l/M < \sqrt{12} \approx 3.46$, $V_{\rm eff}$ is a monotonically
increasing function of the radial coordinate $r$ and a test-particle
coming from infinity is swallowed by the black hole. A comparison
with the case of accreting gas is not straightforward. Moreover,
the initial specific angular momentum tends to increase in our
simulations. Nevertheless, we find a similar behavior for the
accreting gas.

For rotating black hole, the behavior of a corotating disk 
is different from that of counter-rotating one, since spin-orbit 
interactions determine the 
height of the effective potential barrier. For given values of 
$|a_*|$ and $|l|$, the barrier is lower for counterrotating gas, 
which can thus reach the central object even if it has higher angular
momentum.

\begin{figure}
\par
\begin{center}
\includegraphics[height=6cm,angle=0]{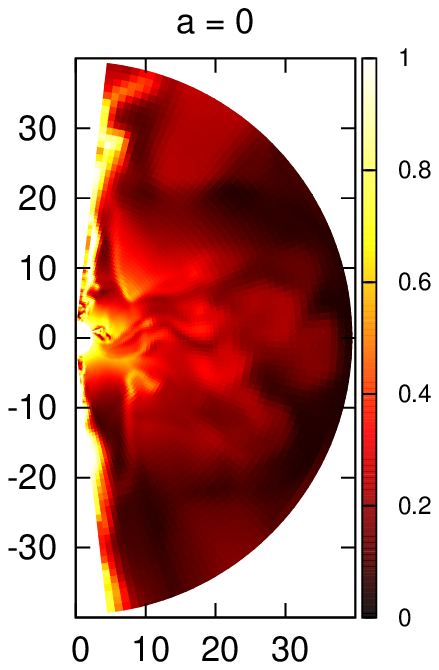}
\includegraphics[height=6cm,angle=0]{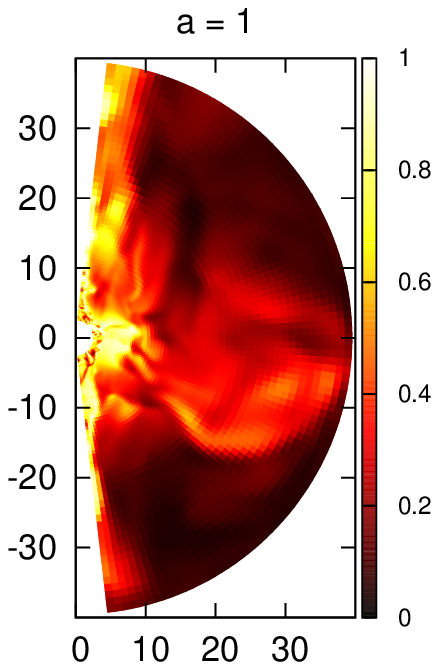}
\includegraphics[height=6cm,angle=0]{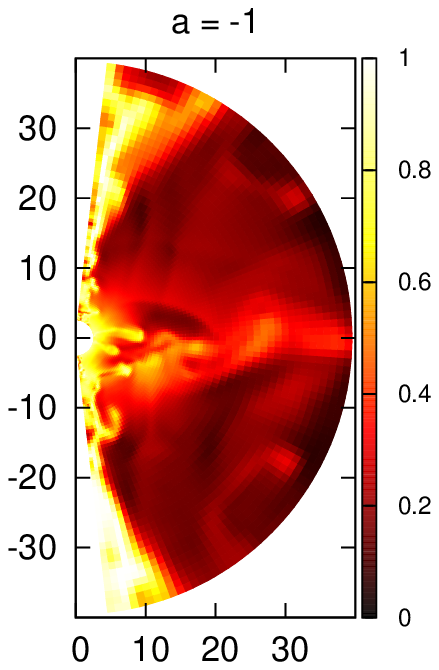}
\end{center}
\par
\vspace{-5mm} 
\caption{Snapshots at $t = 1000$~$M$ of the gas velocity
$v = \sqrt{\gamma_{ij} v^i v^j}$ around black holes with
spin parameter $a_* = 0$ and $\pm 1$. In these simulations,
the accretion process onto Schwarzschild and extreme
Kerr black holes is qualitatively identical.
The unit of length along the axes is $M$.}
\label{f-outflow1}
\end{figure}

In Fig.~\ref{f-outflow1}, 
we show the gas velocity at the time $t = 1000$~$M$
around black holes with $a_* = 0$, $\pm 1$, for $l_0/M = 1.6$. The 
three cases do not present substantial differences. The high velocity
gas in the region along the symmetry axis is a numerical 
artifact and should be ignored. There, the gas density is very low and
the boundary is not fully under control. Fortunately, it does not
singnificantly affect the rest of the evolution.

\subsection{Super-spinars}

As discussed in~\cite{sim2,sim3}, spherically symmetric and adiabatic
accretion onto super-spinars can generate outflows because the 
gravitational field is axisymmetric and, at very small radii, the
force can be either attractive or repulsive. In particular, the radial
gravitational force is strong and repulsive near, but outside, the
equatorial plane; for this reason the matter in the outflows is
ejected perpendicular to the spin of the super-spinar. The effective
gravitational force acting on the falling gas depends, however, even
on the gas angular momentum, which becomes thus an important 
parameter in the dynamics of the accretion process\footnote{Let us
remind the reader that the geodesic equations in Kerr space-time
contain, in addition to the Newtonian centrifugal term, other
contributions proportional to $L_z$, $L_z^2$, and $\mathcal{Q}$,
where $L_z$ is the component of the particle angular momentum 
at infinity parallel to the spin of the massive object and 
$\mathcal{Q}$ is the Carter constant. The latter reduces to 
$L_x^2 + L_y^2$ in the Schwarzschild case $a_* = 0$.}.

\begin{figure}
\par
\begin{center}
\includegraphics[height=6cm,angle=0]{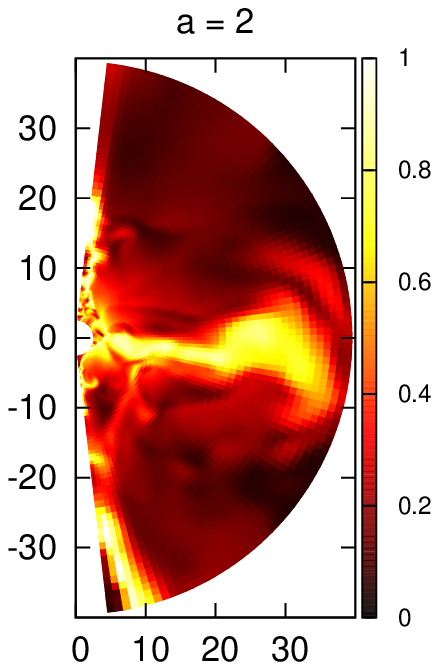}
\includegraphics[height=6cm,angle=0]{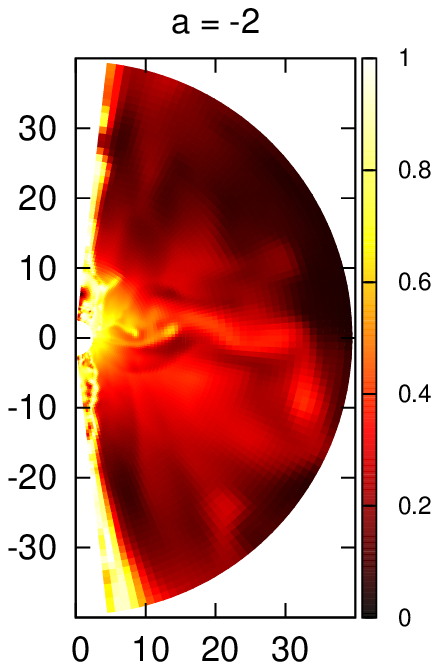}
\includegraphics[height=6cm,angle=0]{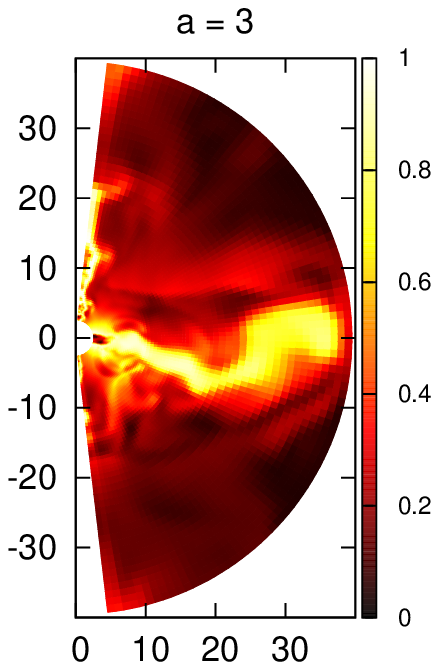}
\includegraphics[height=6cm,angle=0]{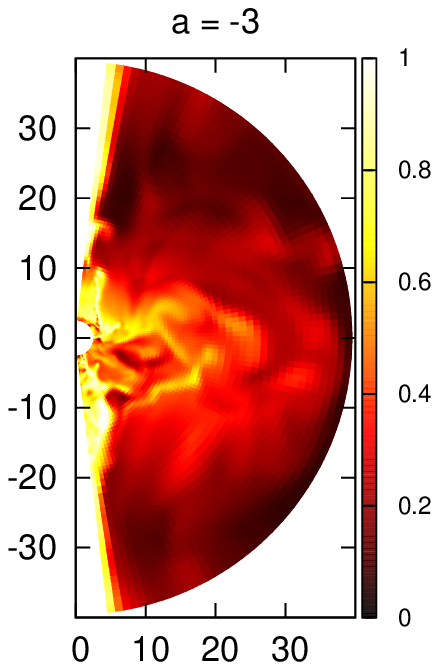}
\end{center}
\par
\vspace{-5mm} 
\caption{Snapshots at $t = 1000$~$M$ of the gas velocity
$v = \sqrt{\gamma_{ij} v^i v^j}$ around super-spinars with
spin parameter $a_* = \pm 2$ and $\pm 3$. In these simulations,
powerful equatorial outflows are present only when the
angular momentum of the accreting matter is parallel to the
spin of the massive object (corotating disk).
The unit of length along the axes is $M$.}
\label{f-outflow2}
\end{figure}

\begin{figure}
\par
\begin{center}
\includegraphics[height=6cm,angle=0]{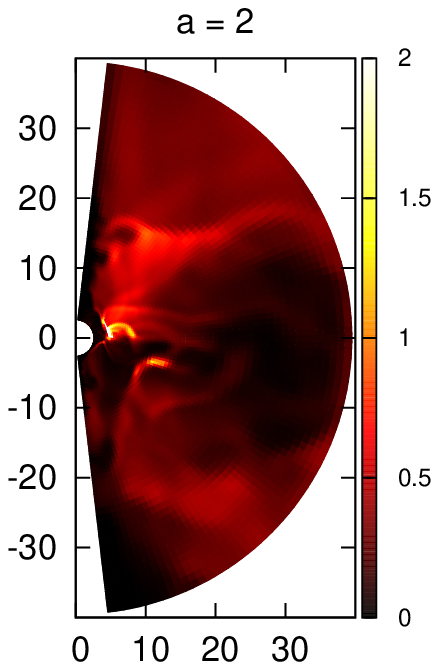}
\includegraphics[height=6cm,angle=0]{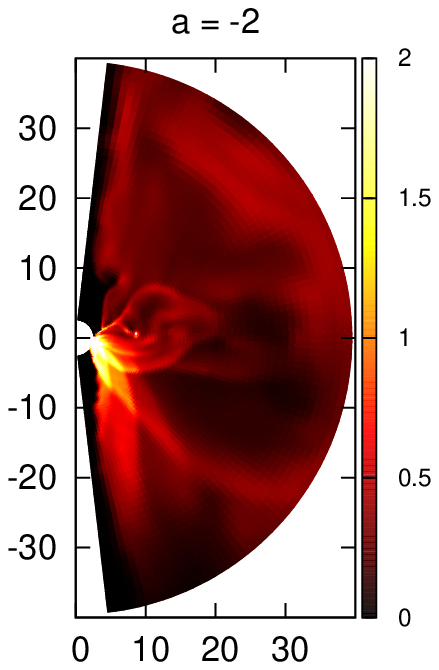}
\includegraphics[height=6cm,angle=0]{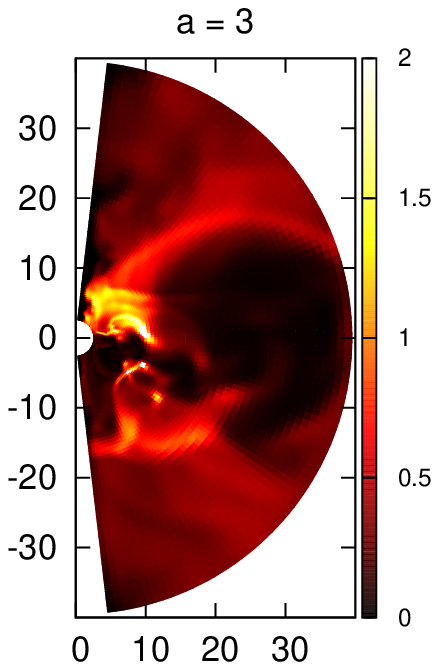}
\includegraphics[height=6cm,angle=0]{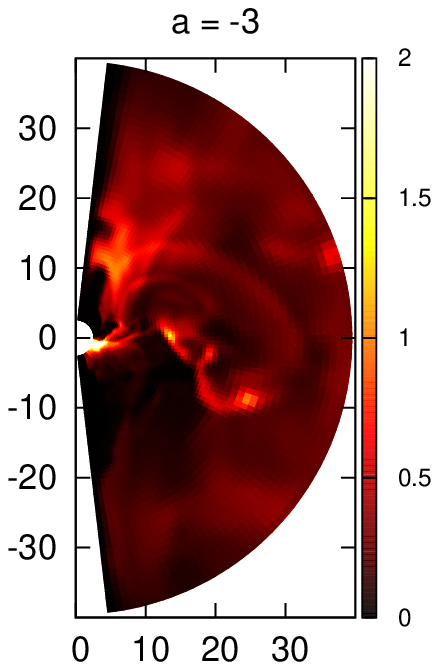}
\end{center}
\par
\vspace{-5mm} 
\caption{Gas density (in arbitrary units) at $t = 1000$~$M$ around
super-spinars with $a_* = \pm 2$ and $\pm 3$. Comparing these plots
with the ones in Fig.~\ref{f-outflow2}, we see that in the
outflows produced in the cases $a_* = 2$ and 3 the gas has low 
density. The unit of length along the axes is $M$.}
\label{f-den2}
\end{figure}

\begin{figure}
\par
\begin{center}
\includegraphics[height=6cm,angle=0]{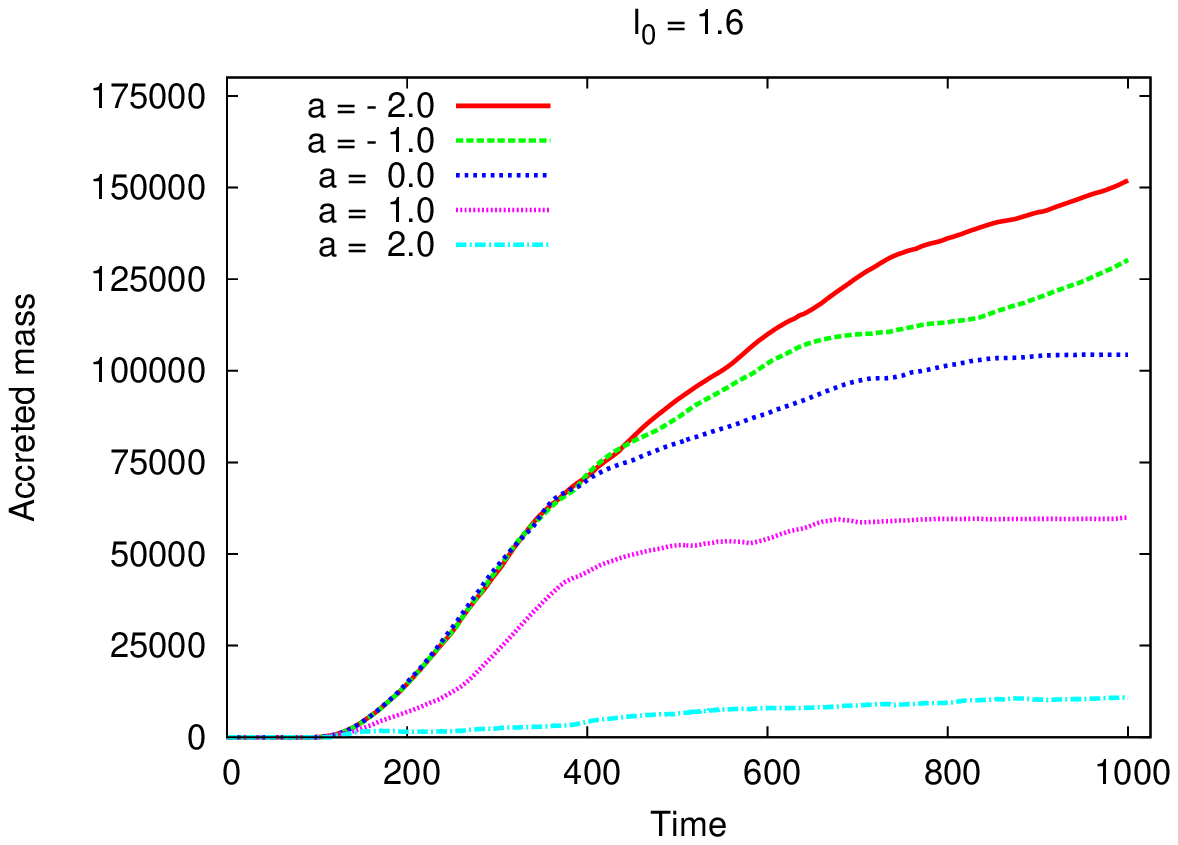} \hspace{.3cm}
\includegraphics[height=6cm,angle=0]{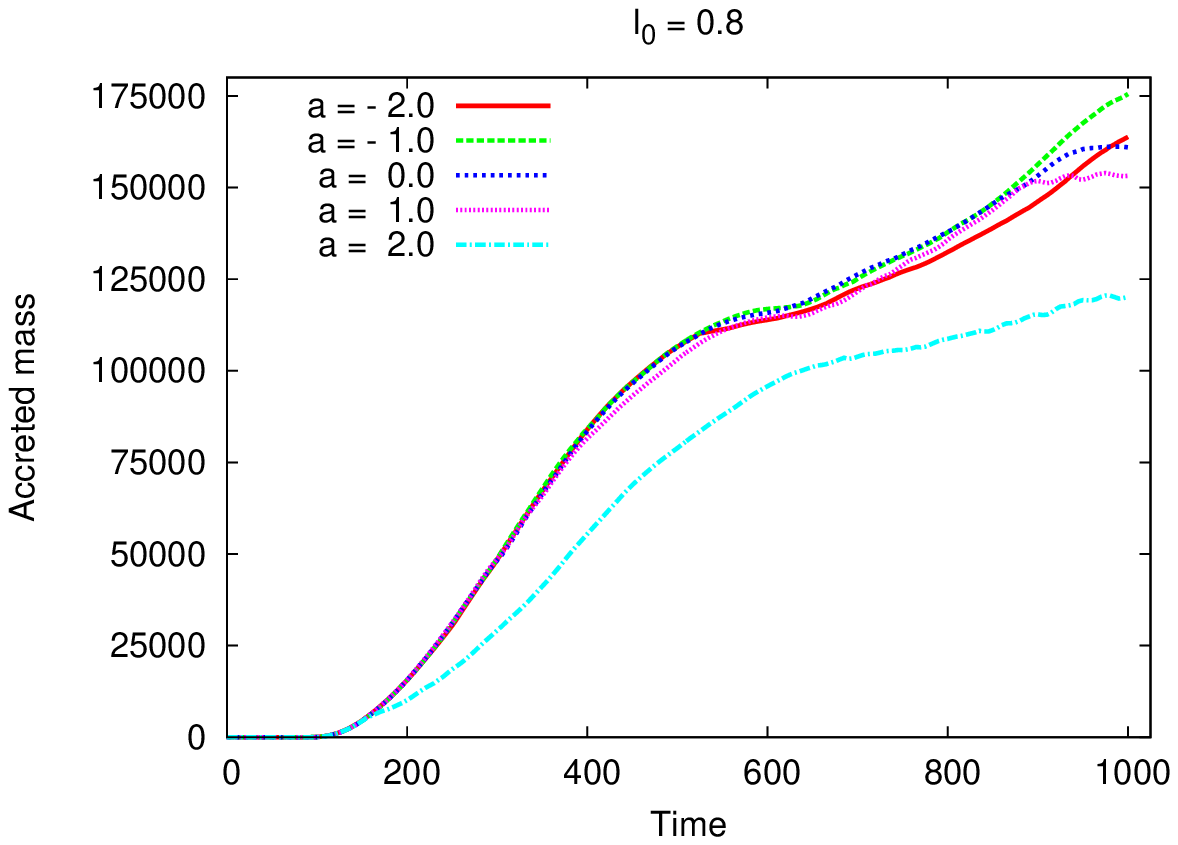}
\end{center}
\par
\vspace{-5mm} 
\caption{Accreted mass as a function of time $t$ of the space 
region inside the radius $r = 5$~$M$ for $a_* = 0$, $\pm 1$ and $\pm 2$.
The injection initial angular momentum per unit mass of the gas
is respectively $l_0 = 1.6$ (left panel) and $l_0 = 0.8$ (right panel).
Accreted mass in arbitrary units; the time $t$ and the specific
angular momentum $l_0$ are in unit $M = 1$.}
\label{f-accrate}
\end{figure}

In Figs.~\ref{f-outflow2} and \ref{f-den2} we show 
respectively the velocity and the density of
the accreting gas around super-spinars with $a_* = \pm 2$ and $\pm 3$,
with our default boundary conditions. In all these cases, the gas 
reaches the central object and there is accretion. However, the
accretion process for corotating and counterrotating disks presents
a significant difference: for $a_* = 2$ and 3, we see equatorial 
outflows characterized by gas with high velocity and low density,
for $a_* = -2$ and $-3$, the accretion looks more like the one onto
black holes, with chaotic and turbulent motion of the gas, but
without equatorial outflows. 
In our earlier work~\cite{sim3}, we found that outflows were
generated for $|a_*| \ge 2.9$ with the same conditions (in particular
$R = 2.5$~$M$), but injecting gas with zero angular momentum.
Fig. \ref{f-den2} shows clearly that a corotating accretion disk favors the
production of outflows, while a counterrotating accretion disk
suppresses it. To be more quantitative, we plot the accreted mass
of gas in the volume $r < 5$~$M$ for black holes and super-spinars,
i.e. the quantity\footnote{Such a quantity is not associated with a
conserved current. From this point of view, the relativistic accreted
mass should be defined as
\be
Q(t) = \int_0^t \int_{r = 5M} \rho u^r \sqrt{-g} 
d\theta d\phi d\tau \, , 
\ee
where $u^r$ is the $r$-component of the gas 4-velocity. However,
here we can use the definition given in Eq.~(\ref{eq-macc}) to 
interpret the data without a significant difference. $M_{acc}(t)$ 
is closer to the Newtonian concept of accreted mass.}
\be\label{eq-macc}
M_{acc}(t) = \int_0^t \int_{r = 5M} \rho v^r \sqrt{\gamma} 
d\theta d\phi d\tau \, .
\ee
The left panel of Fig.~\ref{f-accrate} shows the case $l_0/M = 1.6$:
as the value of the spin parameter decreases, the amount of
gas which can reach the central object is larger. 
For $a_* = 2$, the production
of equatorial outflows reduces considerably the accretion rate
onto the
super-spinar.
For a lower specific angular momentum, we find a trend consistent
with the case of spherically symmetric flow discussed in~\cite{sim3}:
the difference between the cases $a_* = 0$, $\pm 1$ and $-2$, is
apparently very small. When $a_* = 2$ the accretion is much
suppressed, but there are no equatorial outflows (Fig.~\ref{f-accrate},
right panel). Comparing the left and the right panels of 
Fig.~\ref{f-accrate}, we recover the quite obvious result that 
a lower gas angular momentum increases the mass accretion rate,
simply because the effect of the centrifugal force is less important.

Overall, the generation of outflows requires that the
super-spinar is sufficiently fast-rotating and compact, i.e. the
mechanism works for sufficiently high spin parameters $|a_*|$ and
small physical radii $R$. The effect of the gas angular momentum 
is instead more complex. For corotating gas, the height of the
effective potential barrier is not a monotonic function of $a_*$,
but first increases and then decreases as $a_*$ increases. A
higher gas angular momentum favors the production of equatorial 
outflows, but, if it is too high, the gas cannot overcome the
effective potential barrier and thus cannot reach the region close 
to the compact object where outflows can be produced. For counterrotating
disks, the behavior of the accretion process seems to be easier
to predict, because a higher angular momentum increases
the barrier and makes the ejection of matter inefficient.

At this point, one can check numerically the features of the
accretion process by changing $a_*$, $R$, or $l_0$. 
We find that the production of outflows is significantly suppressed
when the disk is counterrotating: for fixed $R = 2.5$~$M$ and
$l_0/M = 1.6$, outflows are possible only for $a_* \lesssim - 5.5$.
Contrarily, outflows are produced very easily for 
corotating disks: for $a_* = 2$ and $l_0/M = 1.6$, we find the
production of powerful outflows for $R \lesssim 4$~$M$.

\subsection{Tilted accretion disk}

In the previous sections, we have adopted the standard
assumption that the accretion disk is symmetric with respect to
the equatorial plane. There are a couple of astrophysical arguments
suggesting that this is the most likely configuration, see e.g.
the footnote in Section~\ref{s-bondi}. However, more recently,
some authors have argued that tilted disks may not be 
rare~\cite{maccarone,fragile}. Numerical simulations of 
tilted accretion disks were presented in~\cite{fra2,fra3}.
It is interesting to study the case with a small but non-zero disk 
misaligned angle with respect to the rotation axis of the compact object. 
In principle, for 2.5D simulations, the choice of the boundary conditions 
at $\theta_{min}$ and $\theta_{max}$ along the symmetry axis may be 
problematic, because now we lose the axial symmetry and therefore 
the axisymmetric boundary conditions are not strictly appropriate. 
Here we consider small tilt angles. We have 
checked that our results are essentially the same for axisymmetric,
periodic, or outflow boundary conditions. In the rest of this section, 
we show the results obtained by imposing outflow boundary conditions,
which have some advantages from the computational point of view.
At the beginning of the simulations, we 
inject gas into the computational domain for 
$\pi/4 - i< \theta < 3\pi/4 - i$, while for $\theta < \pi/4 - i$ and 
$\theta > 3\pi/4 - i$ we impose free-outflow boundary conditions. 
This is our definition of tilt angle $i$.

For systems without equatorial outflows (black holes, super-spinars
with small value of the spin parameter, super-spinar with a
counterrotating accretion disk), we do not see any new feature
associated with the presence of a tilted disk. Here the 
Bardeen-Petterson effect~\cite{b-pet} cannot be seen because the gas 
is too energetic and turbulent, the disk is too thick, and the
mechanism responsible for the transportation of angular momentum
to larger radii is too inefficient. These are all ingredients 
preventing the observation of an accretion disk forced to rotate
in the same plane of the compact object.

\begin{figure}
\par
\begin{center}
\includegraphics[height=6cm,angle=0]{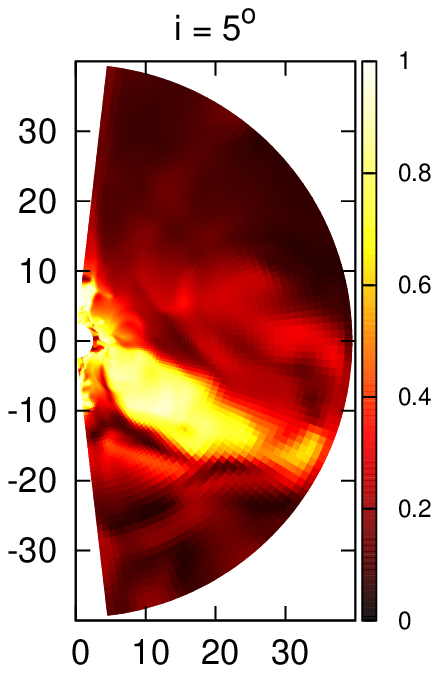}
\includegraphics[height=6cm,angle=0]{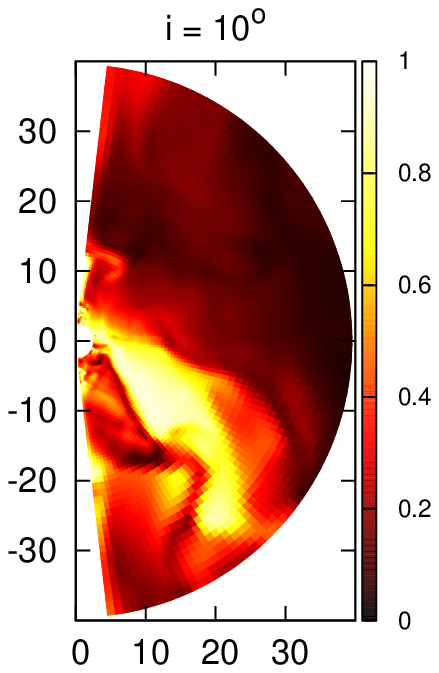}
\includegraphics[height=6cm,angle=0]{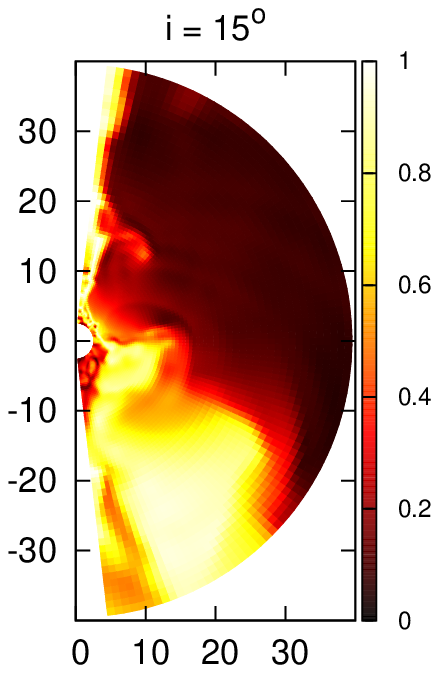}
\end{center}
\par
\vspace{-5mm} 
\caption{Snapshots at $t = 1000$~$M$ of the gas velocity
$v = \sqrt{\gamma_{ij} v^i v^j}$ around a super-spinar with
spin parameter $a_* = 2$. In these simulations, the
corotating disk has a tilt angle $i = 5^\circ$ (left panel),
$i = 10^\circ$ (central panel), and $i = 15^\circ$ (right panel).
The unit of length along the axes is $M$.}
\label{f-tilt-v}
\end{figure}

\begin{figure}
\par
\begin{center}
\includegraphics[height=6cm,angle=0]{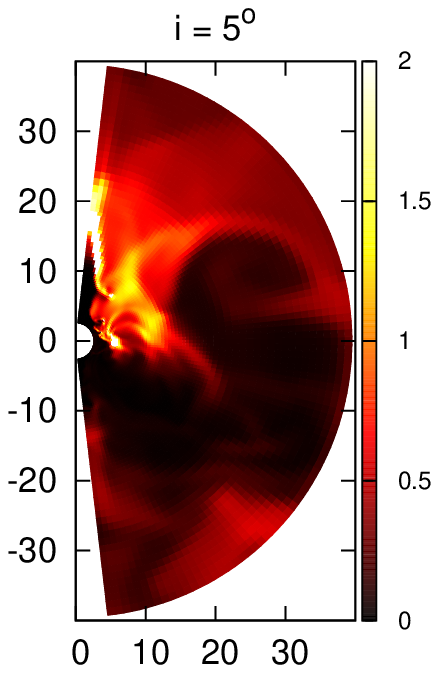}
\includegraphics[height=6cm,angle=0]{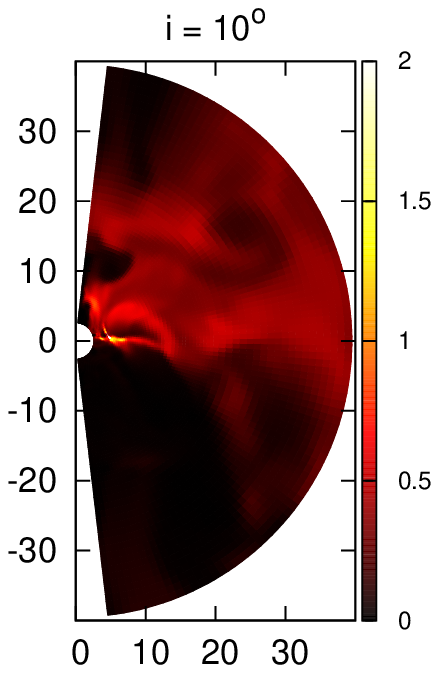}
\includegraphics[height=6cm,angle=0]{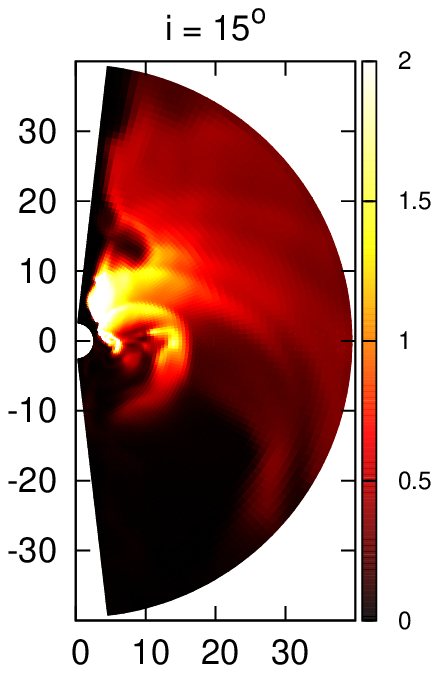}
\end{center}
\par
\vspace{-5mm} 
\caption{Snapshots at $t = 1000$~$M$ of the gas density 
(in arbitrary units) around a super-spinar with
spin parameter $a_* = 2$. In these simulations, the
corotating disk has a tilt angle $i = 5^\circ$ (left panel),
$i = 10^\circ$ (central panel), and $i = 15^\circ$ (right panel).
The unit of length along the axes is $M$.}
\label{f-tilt-rho}
\end{figure}

\begin{figure}
\par
\begin{center}
\includegraphics[height=6cm,angle=0]{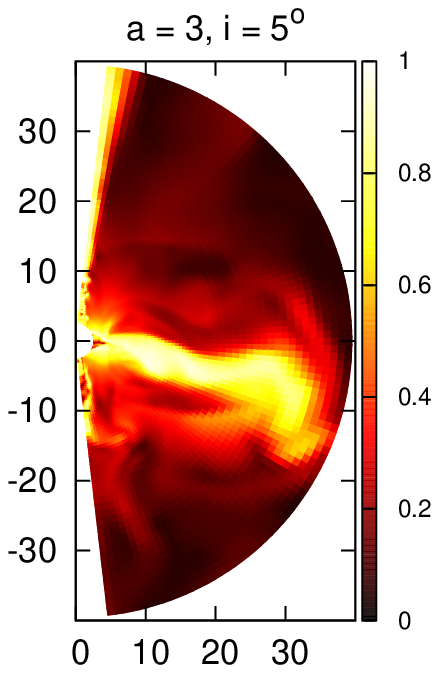}
\includegraphics[height=6cm,angle=0]{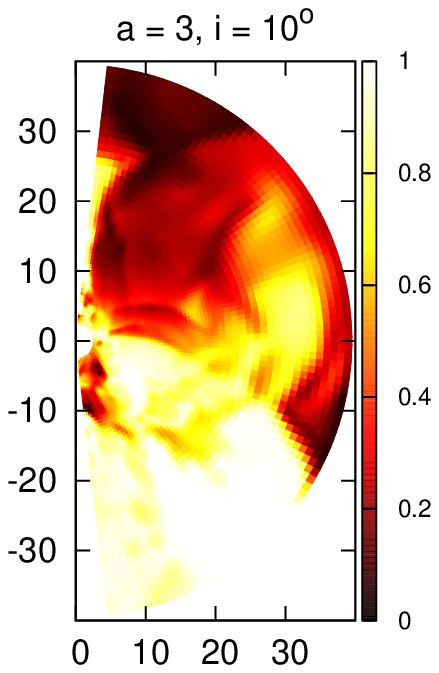}
\includegraphics[height=6cm,angle=0]{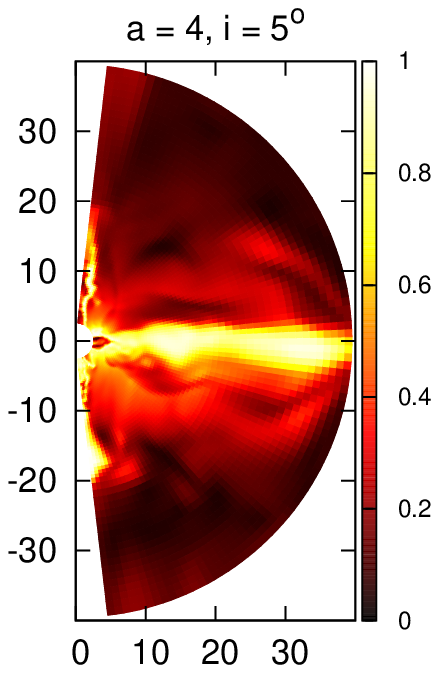}
\end{center}
\par
\vspace{-5mm} 
\caption{Snapshots at $t = 1000$~$M$ of the gas velocity
$v = \sqrt{\gamma_{ij} v^i v^j}$ around a super-spinar with
spin parameter $a_* = 3$ and tilt angle $i = 5^\circ$ 
(left panel), $a_* = 3$ and $i = 10^\circ$ (central panel), 
and $a_* = 4$ and $i = 5^\circ$ (right panel).
The unit of length along the axes is $M$.}
\label{f-tilt-v2}
\end{figure}

The case of super-spinars capable of generating outflows is more
interesting. In Figs.~\ref{f-tilt-v} and \ref{f-tilt-rho}, we 
show respectively the velocity and the density of the gas for a 
tilt angle $i = 5^\circ$, $10^\circ$, and $15^\circ$ and a spin 
parameter $a_* = 2$. In these plots, the spin vector of the
massive object is still along the y-axis. The outflow is not on 
the equatorial plane any more. It is generated close to the 
equatorial plane, but then the gas moves to the side opposite to 
the disk. The reason is that the outflow is not stable and tends 
to follow the easiest way to reach the outer boundary 
where the density of the falling gas is lower.
For higher values of the spin parameter, the outflow is more
energetic. The result is that, for very small tilt angles, the
outflow can still propagate on the equatorial plane; for
more misaligned disks, it propagates out of the equatorial plane
at higher velocities (Fig.~\ref{f-tilt-v2}). At least in the cases
considered here, characterized by small tilt angles, we do not
see any clear difference in the mass accretion rate with respect 
to the configurations with a disk on the equatorial plane.

\section{Conclusions \label{s-conc}}

If current black hole candidates are not Kerr black holes, but compact
bodies made of some kind of exotic matter, the Kerr bound $|a_*| \le 1$
does not hold. The maximum value of the spin parameter may be either
smaller or larger than 1, depending on the properties of these objects. 
The study of the accretion process for different value of the spin
parameter may thus shed light on the nature of the black hole candidates.

In this work we have studied the accretion process in Kerr space-time
with arbitrary value of the spin parameter $a_*$. As already noticed 
in our previous papers~\cite{sim2,sim3}, the accretion process onto 
objects with $|a_*| > 1$ can generate equatorial outflows that can 
be unlikely produced around Kerr black holes. This phenomenon can
be used as observational signature to look for compact objects with
$|a_*| > 1$. Here we have extended 
our previous studies on spherically symmetric accretion to the case
of thick disk accretion, which is a more likely situation for
a binary system, in which the compact object strips matter from
the stellar companion due to tidal effects. Current numerical
studies are not yet capable of simulating geometrically thin disks;
we can however consider geometrically thick disks, which are 
expected when the heat generated by the viscous stress is not 
radiated away efficiently.

Accretion occurs only when the gas can overcome the centrifugal 
barrier, which is determined both by the spin of the compact object 
and by the angular momentum of the gas. For corotating disks, 
the centrifugal barrier is higher; that is, accretion requires 
either a lower gas angular momentum, or a more efficient mechanism 
to lose angular momentum. For counterrotating disks, the centrifugal
barrier is lower and the gas can reach easier the central body.

When $|a_*| > 1$, another effect has to be taken into account. The
gravitational force at small radii and near, but outside, the 
equatorial plane can be strongly repulsive and generate powerful
equatorial outflows. For non-zero gas angular momentum, 
spin-orbits interactions cannot be ignored and the region with 
repulsive gravity is not given by the simple expression 
$r^2 < a^2 \cos^2\theta$, as in the case of spherically symmetric
accretion. Now, equatorial outflows are more easily generated when 
the gas angular momentum is parallel to the spin of the super-spinar, 
while strongly suppressed in the opposite case. The production of 
outflows is determined by the spin parameter $a_*$ and by the 
physical radius of the object $R$, i.e. outflows are possible for 
high values of $a_*$ and low values of $R$. The main result of 
this paper is thus that in a typical binary system, where the
gas angular momentum cannot be neglected, the accretion makes 
outflows possible even for significantly lower $a_*$ and larger 
$R$ for corotating disks, while the production of outflows becomes 
unlikely for counterrotating disks.


\begin{acknowledgments}
This work was supported by World Premier International 
Research Center Initiative (WPI Initiative), MEXT, Japan.
The work of C.B. was partly supported by the JSPS 
Grant-in-Aid for Young Scientists (B) No. 22740147.
The work of N.Y. was partly supported by the JSPS 
Grant-in-Aid for Young Scientists (S) No. 20674003.
\end{acknowledgments}


\end{document}